# NEWSTRADCOIN: A Blockchain Based Privacy Preserving Secure NEWS Trading Network

Anik Islam[1[0000-0002-6725-9805]], Md. Fazlul Kader[2[0000-0002-7139-5965]], Md Mofijul Islam[3[0000-0003-4207-5863]], and Soo Young Shin[4[0000-0002-2526-2395]]

[1, 4] Department of IT Convergence Engineering, Kumoh National Institute of Technology, Gumi, 39177, South Korea
{[1]anik.islam, [4]wdragon}@kumoh.ac.kr
[2] Department of Electrical and Electronic Engineering, University of Chittagong, Bangladesh
f.kader@cu.ac.bd
[3] Department of CSE, University of Dhaka, Dhaka, Bangladesh
akash.cse.du@gmail.com

**Abstract.** In order to stay up to date with world issues and cutting-edge technologies, the newspaper plays a crucial role. However, collecting news is not a very easy task. Currently, news publishers are collecting news from their correspondents through social networks, email, phone call, fax etc. and sometimes they buy news from the agencies. However, the existing news sharing networks may not provide security for data integrity and any third party may obstruct the regular flow of news sharing. Moreover, the existing news schemes are very vulnerable in case of disclosing the identity. Therefore, a universal platform is needed in the era of globalization where anyone can share and trade news from anywhere in the world securely, without the interference of third-party, and without disclosing the identity of an individual. Recently, blockchain has gained popularity because of its security mechanism over data, identity, etc. Blockchain enables a distributed way of managing transactions where each participant of the network holds the same copy of the transactions. Therefore, with the help of pseudonymity, fault-tolerance, immutability and the distributed structure of blockchain, a scheme (termed as NEWSTRADCOIN) is presented in this paper in which not only news can be shared securely but also anyone can earn money by selling news. The proposed NEWSTRADCOIN can provide a universal platform where publishers can directly obtain news from news-gatherers in a secure way by maintaining data integrity, without experiencing the interference of a third-party, and without disclosing the identity of the news gatherer and publishers.

**Keywords:** Blockchain, News Trading, Privacy, Smart Contract.

## 1 Introduction

Blockchain is a distributed database that is replicated and shared among the peers of a network. The blockchain concept was first introduced by Satoshi Nakamoto, a person or a group who used this name as a pseudonym [1]. Any individual from the network can check and verify the ledgers [2]. Valid transactions are stored in the block and each block holds the hash of the previous block. Thus, the process of creating a chain of blocks is called blockchain. However, in order to append data in blockchain, data experiences distributed consensus mechanism and data in blockchain can only be appended and it does not support data erasing [3]. Blockchain can be classified into two ways based on the accessibility of users, such as (1) public blockchain, and (2) private blockchain [4]. In public blockchain, anyone can add data and anyone can participate in the consensus mechanism. On the contrary, in private blockchain, only a predefined list of users can access the network and only owner can participate in the consensus process. However, a mechanism is introduced in blockchain named "Smart Contract" in which if the user knows each other, they can interchange information or transaction without the requirement of any third-party trusted authority [5]. Blockchain has solved trust issues in the distributed system because no party can tamper with the data in the blockchain network [6]. In blockchain, a pair of private/public keys is utilised by the participants in order to communicate with blockchain [2]. Participants employ their public as their identity and exercise their private key to sign transactions which are made by them [7]. Moreover, asymmetric cryptography is used in blockchain in order to maintain its security in the network.

Newspapers play a very significant role in keeping individuals up-to-date with the daily events that have already occurred or are going to happen all over the world. The newspaper is an indispensable part of life for some people who are unable to start their morning without reading one. However, news gathering requires a lot of effort. Every newspaper has their own correspondents who work in different areas of focus and try to collect news regarding their area [8]. Big newspaper publishers have offices in different cities and some have offices in different countries as well. Correspondents, who work in these offices, share news reports very quickly via phone or email [9]. However, those publishers, who do not have a large network of news correspondents, have to buy news from news agencies. Moreover, there is a strong possibility that the collected news may experience modification by the third party or any third party may interfere with the independence of the news gatherer. Moreover, any third party may hinder the process of sharing the collected news. Therefore, news collection from those countries; that are engaged in civil war, massacres or surrounded by terrorists, is very risky and sometimes, it is near impossible for foreign countries to obtain news on the actual situation. Sometimes, news gatherers have to collect news by risking their lives in these countries [10] and sometimes in their own countries [11-13]. A global scheme is required where news gatherers can share their news, while maintaining its integrity, safely and boldly. In addition, news publishers can collect news directly from the correspondent instead of going to the agencies without facing any obstacles.

A cloud-based news sharing scheme is introduced in [14] and another user-generated content based news gathering system is proposed in [15]. Both of them have proposed news sharing schemes only for emergencies. Their proposed scheme cannot solve the aforementioned issues like data integrity, secure data transfer, and pseudonymity. Moreover, the proposed scheme in [14] is centralized, which may pose a threat of third party interference. None of these studies considered blockchain in their proposed scheme. However, the pseudonymity and immutability features of blockchain can help news-gatherers share news fearlessly. Moreover, because of the distributed structure of blockchain, anyone can collect and share news from anywhere without paying a third party. Therefore, by considering the above features of blockchain; in this paper, a blockchain based news trading scheme (termed as NEWSTRADCOIN) is proposed which user can share news securely and also earn money from these news; a topic which has not been explored yet to the best of our knowledge. The major contributions of this paper are compiled as follows.

- The proposed scheme ensures data integrity, no third party interference, and the pseudonymity of news sellers so that they can gather and trade news dauntlessly.
- A new file sharing scheme is proposed where sellers can share news files directly with buyers without disclosing their identity.
- An optimized cost function algorithm is applied so that sellers can share news file in the cheapest way.

The remaining sections of this paper are organized as follows: Section 2 illustrates the system model of NEWSTRADCOIN. The different components of NEWSTRADCOIN are also discussed in this section. In Section 3, different characteristic of NEWSTRADCOIN are discussed in details. A performance comparison among NEWSTRADCOIN and others existing model is demonstrated in Section 4. Finally, Section 5 draws a conclusion from this paper with future research directions.

## 2 Proposed NEWSTRADCOIN

A news trading scheme is devised which uses blockchain concept in order to make it distributed and secured while maintaining the pseudonymity of the news seller. The proposed NEWSTRADCOIN is a new way of trading news. It has the following features:

- News gatherers can sell news dauntlessly by maintaining their pseudonymity.
- News can be shared using a distributed system so that business can avoid system failure. A user does not have to be bound to a centralized authority, and the system can be cheaper.
- News can be shared in a secure way so that no one can cheat each other and news can be tamperproof.
- Create a marketplace where anyone can make money by selling news or working as a file miner.

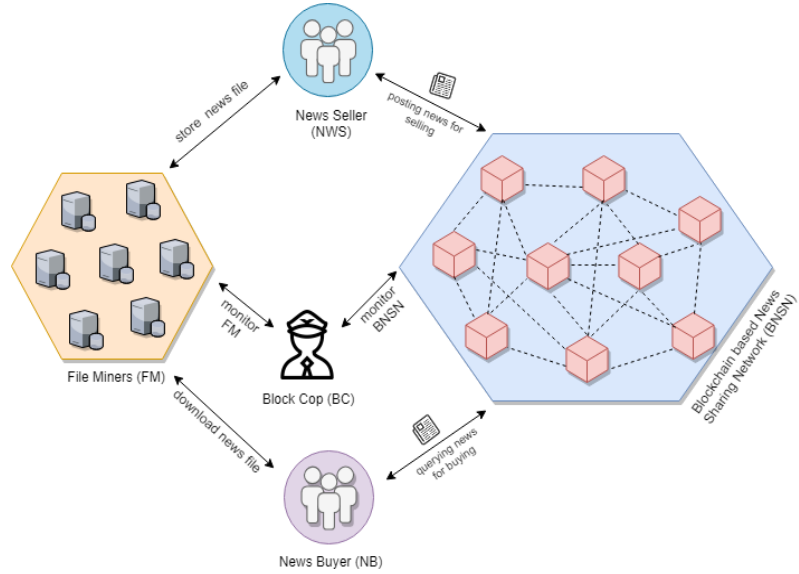

**Fig. 1.** System model of the proposed NEWSTRADCOIN.

The proposed system model is provided in Fig. 1. In the proposed scheme, sellers add their news in blockchain with a price. A secure channel between the user and NEWSTRADCOIN is assumed. Before buying news, buyers perform queries in the blockchain and check the news. If the buyers prefer any news, they can buy the news through NEWSTRADCOIN. When sellers receive money from buyers through the proper channel, sellers send news documents to buyers using NEWSTRADCOIN. The five major components of NEWSTRADCOIN are news seller (NWS), news buyer (NB), news miners (NM), file miners (FM), and block cop (BC). The description of these entities are provided below:

- **News Seller (NWS)** - In this entity, users are the sellers of news. They collect and sell news via NEWSTRADCOIN.
- **News Buyer (NB)** - In this entity, users are the buyers. They can search news to buy based on their preferred categories. Buyers can also act as a seller and seller can act as a buyer and vice versa.
- **News Miners (NM)** - In this entity, miners mine news which is posted from NWS. After validating block by its hash, they assign news blocks in the block-chain network. In blockchain, there is a ledger which contains some part of the news (headline, some description etc.) and it is distributed to the participant nodes. NWS and NB can also lend their resources for mining.
- **File Miners (FM)** - In this entity, miners mine files which are shared from NWS. When users from NB buy news from NWS, NWS obtains assistance from FM to share that news file with the corresponding NB by maintaining file security and the sellers' pseudonymity.

- **Block Cop (BC)** - This entity provides protection against cheating. Suppose, a user sells a news and takes money from the buyer. If the seller does not send news files to the buyer then BC investigates the case using transaction history which is stored in the blockchain ledgers. Finally, BC imposes high penalties to the wrongdoers.

## 3  Features of NEWSTRADCOIN

### 3.1  Distributed and Secure

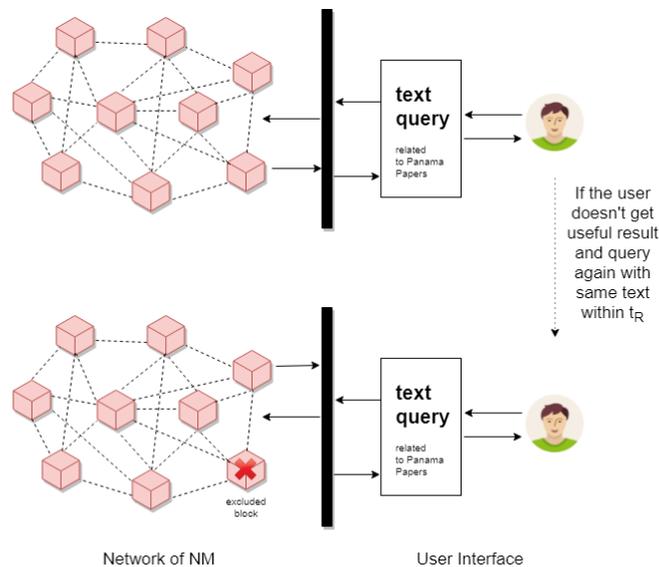

**Fig. 2.** Selection of miner for performing query.

NEWSTRADCOIN is a distributed marketplace where there is no central authority or any third party to maintain transactions. In NEWSTRADCOIN, when a user from NWS posts news to sale, it experiences multiple steps before adding it to blockchain. First, NM validates the transaction based on the hash generated from the block. It also checks for duplicate entries. Multiple miners engage in this process. The miner who validates the block first, obtains money as a reward, as shown in Fig. 2. However, those who do not obtain a reward, store the news for future queries. Moreover, the privacy of the news is very important. It is the sellers' decision how much or to what extent they want to expose to buyers. Moreover, blockchain maintains the immutability of the data using its technology. Hence, news tampering is not possible in NEWSTRADCOIN. NM not only mines news blocks but also gathers data for the queries which are executed by buyers. When buyers make queries in the system in order to find expected news,

NM engages in the data gathering process. The miner who provides the result first obtains the reward. If a buyer does not obtain any useful data, the buyer again makes queries using the same text in the system within a certain time range $t_R$, as shown in Algorithm 1. NEWSTRADCOIN maintains the security of data with the help of blockchain. Once the news is posted on the network, no one can alter the content. If anyone tries to change the news, that change has to be validated by the other members of the network. However, after purchasing the news, the original content is shared in a secure channel with the assistance of encryption that not only ensures the integrity of the data but also precludes the interference of a third party over the content.

### 3.2 Secure File Transfer

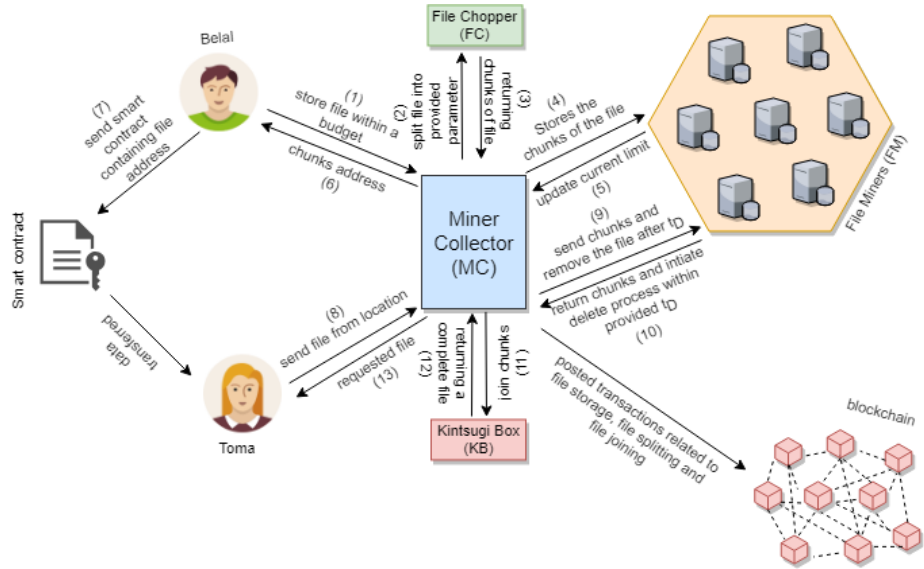

**Fig. 3.** Proposed secure file sharing model.

The basic structure of blockchain does not support storing files because it can increase the size of a block which requires a lot of space. Blockchain generally contains the transaction history or text-based data. Hence, NEWSTRADCOIN does not store files in the NM. Instead, it uses another network of miners who rent their resources to hold news files. If a user wants to share a file with a buyer, that the buyer has already purchased, then that user can easily transfer the corresponding news file using the proposed file sharing scheme, as shown in Fig. 3. Suppose, Toma purchases news from Belal. Then, Belal sends the file to Toma as described in Fig. 3. Firstly, Belal uploads the file to the system with a budget to store the file in FM. Miner collector (MC) receives that request from Belal. The main tasks of MC are as follows: (1) holding the miners' information with their latest space limit and costs per unit of space, (2) storing files in miners' workspace and retrieving files from miners', (3) handling user requests of storing and retrieving files, (4) splitting files with the help of the file chopper (FC) which is a module for splitting files, (5) joining files with the help of Kintsugi box (KB) which is a module for joining a file that has been splitted, and (6) posting transactions history to blockchain. Secondly, MC splits files into chunks so that MC can store those chunks in different miners. As MC contains the information of the miners, including their storage space limit and file storing charge, MC calculates how many chunks are needed for storing. When a user sends a request to store files, he/she sends a file along with his/her budget. Before splitting the file, MC first calculates the unit budget price. Let b is the unit budget price. If S is the file size and B is the budget then

$$b = \frac{B}{S} \quad (1)$$

After that, MC selects FMs based on the cost per unit space calculated by $fee(x)$ and available space, as shown in Algorithm 2. According to Algorithm 2, MC first calculate $b$ using Eq. (1) which is the budget per byte. Here, the system is considering the byte as a file unit and miners' charge is also per byte. Subsequently, MC goes through all of the FMs and checks the following: (i) whether the fee provided by anyone is equal or less than $b$ and (ii) whether they have free space or not. If MC finds any FM, then MC adds that FM to the list of selected candidates SFM for storing files. There is a threshold $T$ to show how many miners can participate in the file storing process.

**Algorithm 1.** Query execution

$t_R$ : time range of re-query.
$t_c$ : current timestamp.
**if** *user is not satisfied and $t_c \leq t_R$* **then**
    exclude the current miner.
    allow user to re-query using previous coin.
**else**
    user have to stay with result.
**end**

**Algorithm 2.** Selection of FMs

$S$ : size of the shared file.
$B$ : sender's budget.
$LFM$ : list of $FMs$.
$SFM$ : list of selected $FMs$.
$T$ : threshold for selecting $FMs$.
$b \leftarrow \frac{B}{S}$.
**while** $x \in LFM$ **do**
    **if** $fee(x) \leq b$ and $space(x) > 0$ **then**
        $Append\_To\_SFM(x)$.
        **if** $length(SFM) == T$ **then**
            break.
        **end**
    **end**
**end**

If SFM reaches $T$ then MC stops looking for miners and starts selecting how many miners can participate in the final process. In the final selection process, MC first sorts the selected FMs into ascending order based on the fee. Let SFM be the list of selected FMs. Therefore, the list of sorted FMs in ascending order is denoted by $SSFM_{FEE}$, where the subscript $FEE$ represents the cost per unit space. After that, it continuously selects a FM and checks for space, as shown in Algorithm 3. When MC generates the final list of FMs, it sends that list to FC. As FC receives a list of FMs, it starts its chopping process, as shown in Algorithm 4. As the list is already sorted based on the fee, the cheaper miner gets priority for renting space. FC stores FM with an allotted size. There is a chance that the last chunk is smaller than the space of FM. Therefore, for the last chunk, space deduction is not needed. When the process is finished, FC returns the chunks with FMs and their allotted size. As MC gets a response from FC, MC sends chunks to the selected FMs. Finally, MC responds to Belal with the addresses of chunks. Before uploading a file, Belal encrypts the file by applying an asymmetric key algorithm. Hence, every chunk of the file is also encrypted. Let $F$ be the file that Belal wants to share, $K_{pub}^{rec}$ is the receivers public key. If $F_E$ is the encrypted file then

$$F_E = E_{K_{pub}^{rec}}(F, K_{pub}^{rec})$$

In order to access the file, Toma has to decrypt using her private key. As Belal obtains addresses of chunks from MC, he creates a smart contract containing the tuple is created:

$$\langle K_{pub}^{rec}, \{FC_{a_0}, FC_{a_1}, FC_{a_2}, \ldots, FC_{a_n}\}, R_a \rangle$$

Here, $R_a$ is the receiver address and $FC_{a_n}$ is the address of $n^{th}$ $(n = 0,1,2,\ldots)$ chunk of the file where $n$ is the total number of chunks address.

**Algorithm 3.** Selection of final FMs

$S$ : size of the shared file.
$rs$ : remaining space.
$SFM$ : list of selected $FMs$.
$FSFM$ : final list of sorted $FMs$.
$SSFM \longleftarrow SORT(SFM, ASC, FEE)$.
$rs \longleftarrow S$.
**while** $x \in SSFM$ **do**
　　$rs \longleftarrow (rs - space(x))$.
　　$Append\_To\_FSFM(x)$.
　　**if** $rs \leq 0$ **then**
　　　　$break$.
　　**end**
**end**

**Algorithm 4.** Selection of space for chunks

$S$ : size of the shared file.
$rs$ : remaining space.
$FSFM$ : final list of sorted $FMs$.
$FMS$ : list of $FMs$ with allotted size.
$rs \longleftarrow S$.
**while** $x \in FSFM$ **do**
　　**if** $(rs - space(x)) < 0$ **then**
　　　　$Append\_To\_FMS(x, rs)$.
　　**else**
　　　　$rs \longleftarrow (rs - space(x))$.
　　　　$Append\_To\_FMS(x, space(x))$.
　　**end**
**end**

When Belal finishes creating the smart contract, it is sent to Toma. As Toma obtains the smart contract from Belal, the addresses of chunks are sent to MC to collect the chunks and combine the chunks into a file. When MC obtains the address, MC collects chunks from the addresses and initiates a timer in the FM to delete chunks after a certain amount of time $t_D$. After passing $t_D$, FM deletes chunks from their storage. However, MC sends the chunks to Kintsugi Box (KB). The main task of KB is to join the chunks into a file. When KB finishes joining, KB returns the file to MC. After that, MC returns the file to Toma. Let $K_{pri}^{rec}$ be the private key of the receiver. Finally, Toma decrypts the file. If $F_D$ is the decrypted file then:

$$F_D = D_{K_{pri}^{rec}}(F_E, K_{pri}^{rec})$$

### 3.3　Pseudonymous

One of the major benefits of this proposed scheme is users' pseudonymity. Instead of giving personal information, the proposed scheme uses blockchain's pseudonymity techniques. In blockchain, asymmetric encryption is adopted in order to maintain its security. However, public keys are used as a signature in order to identify a user. Instead of linking real-life information to transactions, user signature is linked with the transaction. Every user in this system has a unique signature which they use as an identity. NEWSTRADCOIN generates a signature using a timestamp $t$, nonce $n$, a random text $T$ from a user, and a salt hash $h$. Let $S = Sig(t, n, T, h)$ is the signature of a user. The system generates $S$ for a new buyer/seller in order to maintain transactions without revealing their identity.

### 3.4　Trustable

NEWSTRADCOIN is a trustable marketplace. NEWSTRADCOIN contains the fraud detection mechanism on the NM and FM. If a buyer reports a case against a seller that the seller did not send the file after getting money, then BC investigates the case. When buyer transfer money to the seller, there is a transaction history in the blockchain. When a seller sends a smart contract to a buyer and that buyer joins and decrypts that file in NEWSTRADCOIN, there is two transaction history in the blockchain. BC goes through the transaction history and checks the abnormality of transactions. If BC finds anything fishy, BC notifies the seller to solve the issue within a deadline. If the seller does not solve the issue with the buyer then BC imposes a high penalty on the seller and return the money back to the buyer. Moreover, if any FM fails to provide service after getting the fee from hoster, BC also imposes a penalty on FM. This creates trust in the system and prevents peer to involve any illegal activities.

### 3.5　Profitable

NEWSTRADCOIN provides a profitable marketplace. There is a lot of opportunity for earning money. First of all, anyone can collect and sell with very few charges. As sellers are pseudonymous, they can sell news dauntlessly. Moreover, NEWSTRADCOIN contains earning options for non-sellers. In NM, anyone can earn money by mining and generating data based on a query from a buyer. In FM, anyone can earn money by renting their storage space. Sellers can use their space to transfer files to buyers which helps them to remain pseudonymous. Moreover, NEWSTRADCOIN selects cheaper space providers for sellers within a seller's budget. Let $T_p$ be the total price, if $n$ is the total number of selected FMs and $fm$ is the file miner then:

$$T_p = \sum_{i=0}^{n}(as_i \times fee(fm_i))$$

where $T_p \leq B$. Here, $B$ is the budget of a seller and $as_i$ is the allotted space for the $i^{th}$ $fm$. The proposed system is able to optimize the price so that seller can store their file for a cheap price.

## 4  Performance Comparison

The performance comparison of NEWSTRADCOIN with Kumar et al. [14] and Zhang et al. [15] is demonstrated in Table 1. Here, "Mess sharing" means any user can share news which is supported by Kumar et al. [14], Zhang et al. [15] and NEWSTRADCOIN, pseudonymity means anyone can share news without disclosing their identity which is only supported by NEWSTRADCOIN, immutability means no one can alter the news after sharing which is only supported by NEWSTRADCOIN, secure transfer means the transfer of content in a secure channel which is only supported by NEWSTRADCOIN, and distributed means there is no central authority in the system which is supported by both Zhang et al. [15] and NEWSTRADCOIN.

**Table 1.** Performance comparison between NEWSTRADCOIN and existing models

| Features | Proposed Schemes | | |
|---|---|---|---|
| | Kumar et al. [14] | Zhang et al. [15] | NEWSTRADCOIN |
| Mess sharing | ✓ | ✓ | ✓ |
| Pseudonymity | ✗ | ✗ | ✓ |
| Immutability | ✗ | ✗ | ✓ |
| Secure transfer | ✗ | ✗ | ✓ |
| Distributed | ✗ | ✓ | ✓ |

## 5  Conclusions and Future Work

In this paper, we proposed a secure and distributed news trading scheme exploiting the pseudonymity, immutability and distributed mechanism of blockchain. In the proposed scheme, news-gatherers sell news to buyers while maintaining data integrity and without exposing their identity. Moreover, after the deal, sellers can maintain this pseudonymity while sharing necessary documents with buyers in the cheapest way. Therefore, our proposed scheme can easily be adopted by any news collection chain as a private network for collecting news, with little or no modification. Information regarding implementation along with detailed numerical results is kept for the extension of this paper. Furthermore, the reputation system of buyers and sellers for managing the authenticity of content and deals, and a detailed plan for managing the wallet for easy payment can be incorporated, which can be a subject for future studies.

## Acknowledgement

This work was supported by Priority Research Centers Program through the National Research Foundation of Korea (NRF) funded by the Ministry of Education, Science and Technology (2018R1A6A1A03024003).